\documentclass[aps,prl,10pt,
twocolumn,superscriptaddress,showpacs,shortbibliography]{revtex4-1}

\usepackage{graphicx}
\usepackage{dcolumn}
\usepackage{bm}
\usepackage{caption}
\usepackage{subcaption}
\usepackage[utf8]{inputenc}
\usepackage[T1]{fontenc}
\usepackage{mathptmx}
\usepackage{hyperref}
\usepackage{cleveref}
\usepackage{comment}
\usepackage{float}
\usepackage{color,soul}
\newcommand{\cblack}{\color{black} }

\graphicspath{{./images_fin/}}
\begin{document}
	\pagenumbering{roman} \normalsize
	\pagenumbering{arabic}
	\title{Equation of State of a Strongly Interacting many-Boson System from an Effective Interaction}
	\author{Hilla\ De-Leon}
	\affiliation{INFN-TIFPA Trento Institute of Fundamental Physics and Applications, Via Sommarive, 14, 38123 Povo TN, Italy}
	\affiliation{
		European Centre for Theoretical Studies in Nuclear Physics and Related Areas (ECT*),
		Strada delle Tabarelle 286, I-38123 Villazzano (TN), Italy}
	\author{Francesco Pederiva}

	\affiliation{INFN-TIFPA Trento Institute of Fundamental Physics and Applications, Via Sommarive, 14, 38123 Povo TN, Italy}
		\affiliation{Dipartimento di Fisica, University of Trento, via Sommarive 14, I–38123, Povo, Trento, Italy}
		\email [Email:~]{francesco.pederiva@unitn.it}
	\begin{abstract}
		A contact  potential describing an effective interaction between atomic $^4$He reproducing the results obtained with the HFDHE2 potential by Aziz et al. \cite{doi:10.1063/1.438007} is employed to study the resulting equation of state by means of Quantum Monte Carlo calculations.  \cblack The energy per particle and the pair distribution functions were investigated as a function of the ultraviolet cutoff $\Lambda$. The results suggest that not only the mean field properties of the system, such as the energy and the saturation density, are correctly reproduced, but also very microscopic quantities such as the pair distribution function are seen to converge towards the exact results when extrapolating for $\Lambda\rightarrow\infty$.
	\end{abstract}
	\maketitle
	

The use of effective interactions to simplify the approach to the study of otherwise extremely complex systems is widespread in physics. The interaction between atoms and molecules is often described in terms of potentials that can be constructed starting from an accurate study of the underlying Coulomb interactions. In nuclear physics, rather than solving the problem at the Quantum ChromoDynamics (QCD), one of the main goals of modern nuclear physics is to obtain a precise and robust theory leading from QCD to nuclear observables, such as nuclear masses, electroweak interactions, etc. In general, a direct calculation of physical observables for such many-body systems is not trivial due to the non-perturbative character of the underlying interactions. However, in the last decades, a general approach for studying low energy reactions, that we will refer to as Effective Field Theory (EFT), has been developed that aims to establish a perturbative hierarchy in the estimation of the observables and to provide a systematic way to estimate the error related to the truncation of the expansion at a given order. The condition for describing a physical process using EFT is that the typical exchanged momentum scale $Q$ is small compared to some large typical energy cutoff $\Lambda_{\rm cut}$. For example, in nuclear physics, the binding energy of deuterium and the related momentum exchanged between the proton and the neutron (few MeV) is small compared to a cutoff determined by the energy needed to create a virtual pion in the interaction, i.e., $\sim$140 MeV. In this case, the most general Lagrangian (or Hamiltonian), including only the relevant degrees of freedom below $\Lambda_{\rm cut}$, and preserving all symmetries of the fundamental interaction, is constructed, while heavier excitations are integrated out of the action. Thus, one obtains an expression of non-renormalizable interactions, which can be organized as a power series in $\frac{Q}{\Lambda_{\rm cut}}$, i.e., such as the nuclear observables can be written as $O_\Lambda=O_0+\frac{C_1}{\Lambda_{ \rm cut}}+\frac{C_2}{\Lambda_{cut}^2}+ ...$ ~\cite{few, kaplan1, KSW1998_a, KSW1998_b, KSW_c}. This is the so-called
	"pionless" version of nuclear EFT. In fact, the momentum scale suggests that the only degrees of freedom needed to determine the structure of a nucleus are nucleons interacting, at leading order, through contact interactions only. 
When dealing with atomic or molecular systems, the situation is to some extent similar. The underlying fundamental interaction describing the occurring chemical and physical properties is the Coulomb force exerted among electrons, and between electrons and ions. However, for very low energy processes like those described in terms of Van der Waals or dipolar forces, it is convenient to consider atoms as elementary degrees of freedom and utilize some effective potentials, usually presenting a hardcore and some attractive part, which are typically fitted on experimental phase shifts or results of quantum chemistry calculations.

There have been several attempts of employing in condensed matter systems a perturbative approach similar to that used in nuclear physics (Refs.~\cite{Bazak:2016wxm,Platter:2004he,PhysRevA.104.L030801} for example), in particular for the case of cold, dilute atomic gases and for atomic $^4$He condensed systems. In the latter case, the approach is justified by the fact that this bosonic system has two length scales, $a_2$ and $r_{Vdw}$, with scale separation between the two, making the two-body $^4$He system ideal for an effective potential. For the $^4$He atomic system, the two-body scattering length, $a_2 \approx 170.9 a_0$ is much larger than its
van der Waals radius, $r_{VdW} \approx 9.5 a_0$ with $a_0$ is Bohr radius. Hence, for the N-body $^4$He system, there is a large separation between the length scale of interest ($a_2$) and the length scale of the underlying dynamics ($r_{Vdw}$). 
Native EFT potentials are normally believed to make sense in the context of few-body and/or diluted systems. In Bosons, contact interactions have also been proved to give sensible results for relatively large systems. However, there is an aspect that needs to be explored and that is extremely relevant in the context of nuclear physics. It is known that nucleons, as well as atoms, need to be described by a strongly repulsive short-range force. In a bulk of relatively high density, such short-range forces are crucial to determine the correct properties of the system and, ultimately, the Equation of State. Whether or not a contact interaction and, in general, a perturbative EFT approach is able to reasonably reproduce the behavior of homogeneous matter, especially at densities above saturation, is a perfectly sensible question. Recently Kievsky et al. \cite{PhysRevA.104.L030801} performed an interesting analysis of the extrapolation to the bulk in terms of Density Functional Theory, showing that the extrapolation is somewhat coherent with old GFMC results by Pandhariphande at al. \cite{PhysRevLett.50.1676}. In this letter, we want to investigate the problem by using Quantum Monte Carlo methods to compute the Equation of State of $^4$He near the saturation density, using a Leading Order (LO) effective interaction fitted on small droplets simulated by means of a realistic Aziz He-He potential\cite{PhysRevA.104.L030801}, and then extending the calculations to a bulk at different densities, to observe the evolution with the cutoff defining the effective interaction of observables such as the binding energy per particle and the pair distribution functions. This problem sheds light on at least one of the difficulties that are believed to plague the extrapolation of nuclear forces to describe homogeneous nucleonic matter, stripping out the effects due to the fermionic nature of the nucleons and on the relative spin and angular momentum-dependent nature of any nuclear potential.
	
The Diffusion Monte-Carlo (DMC) algorithm (see e.g. \cite{Carlson:2014vla, RevModPhys.73.33} ) is based on a stochastic implementation of an imaginary time propagator $\exp[-t(H-E_0)/\hbar]$, where $E_0$ is an estimate of the ground state energy, applied to some arbitrary initial state $|\Psi_T\rangle$ to project out the ground state $|\Psi_0\rangle$, provided that $\langle \Psi_0\vert\Psi_T\rangle\neq 0$.
	This method is particularly powerful in the case of many-Boson problems. The specifics of the code we used for implementing DMC follow the standards described in literature. In DMC particular care must be taken in constructing the wavefunction to be used as inital state and as importance function for the walks to be generated. Such function must be a reasonable approximation of the ground state, especially for as concerns the short range analytic behavior, in order to reduce the statistical noise on the results. 
	
 In order to construct our wavefunction we first solve the two-body problem using the bare two-body Hamiltonian:
\begin{equation}\label{eq_H2_B}
\hat{H}_{\text{2-B}}=-\frac{\hbar^2}{2M_{\text{relative}}} \nabla^2+V_2 (\Lambda)e^{-\boldsymbol{r}^2\cdot\Lambda^2}~.
\end{equation} 
$\Lambda$ plays the role of an ultraviolet cutoff that regulates the two-body potential, $V_2(\Lambda)$ is a so called Low Energy Constant (LEC), and $\textbf{r}$ is the relative distance between two $^4$He atoms.
The two-body wave-function is given by the solution of:
\begin{equation}\label{eq_E2}
\hat{H}_{\text{2-B}}\psi_{\text{2-B}}(r,\Lambda)=E_2(\Lambda)\psi_{\text{2-B}}(r,\Lambda)=E_2\psi_{\text{2-B}}(r,\Lambda).
\end{equation}
 The last equality points out the fact that in a renormalized theory the binding energy of the two-body system must be independent of $\Lambda$. For such reason, for each $\Lambda$, $V_2(\Lambda)$ is calibrated using \cref{eq_E2,eq_H2_B} assuming $E_2=0.83012$ and
$\frac{\hbar^2}{2M_{\text{relative}}} = 43.281307 K a_0^2$ with $a_0$ the Bohr
radius\cite{Kievsky:2017mjq}. 
This calibration was done by means a direct numerical solution (Chebyshev algorithm \cite{10.1093/comjnl/4.4.318}) of the two-body differential equation for different values of the cutoff ($\Lambda$). 

The next step in writing the equation of state of many-body system is to solve the $N=3$ ($^4$He trimer) Schrodinger equation using the calibration of $V_2(\Lambda)$ found in the dimer as a sum of two-body interactions \cite{Barnea:2013uqa,Kirscher:2015yda}:
\begin{equation}
\label{eq_three_1}
\hat{H}_{\text{3-B}}=
-\sum_{i<j} \frac{\hbar^2}{2M_{\text{cm}}}
\nabla_{ij}^2+V_2 (\Lambda)e^{-r_{ij}^2\cdot\Lambda^2},
\end{equation}
where the indices $i,j=1\dots 3$, and $\nabla^2_{ij}$ refers to the Laplacian with respect of the relative coordinate of the $i,j$ pair.
In contrast to the dimer, here we used the QMC algorithm to solve eq. \ref{eq_three_1}, with the analytical dimer wave-function ($\psi_{\text{2-B}}(\Lambda)$) as $\Psi_T$.
	\begin{figure}[h]\label{fig_E3}
	\begin{subfigure}[b]{0.235\textwidth}
	\includegraphics[width=1\linewidth]{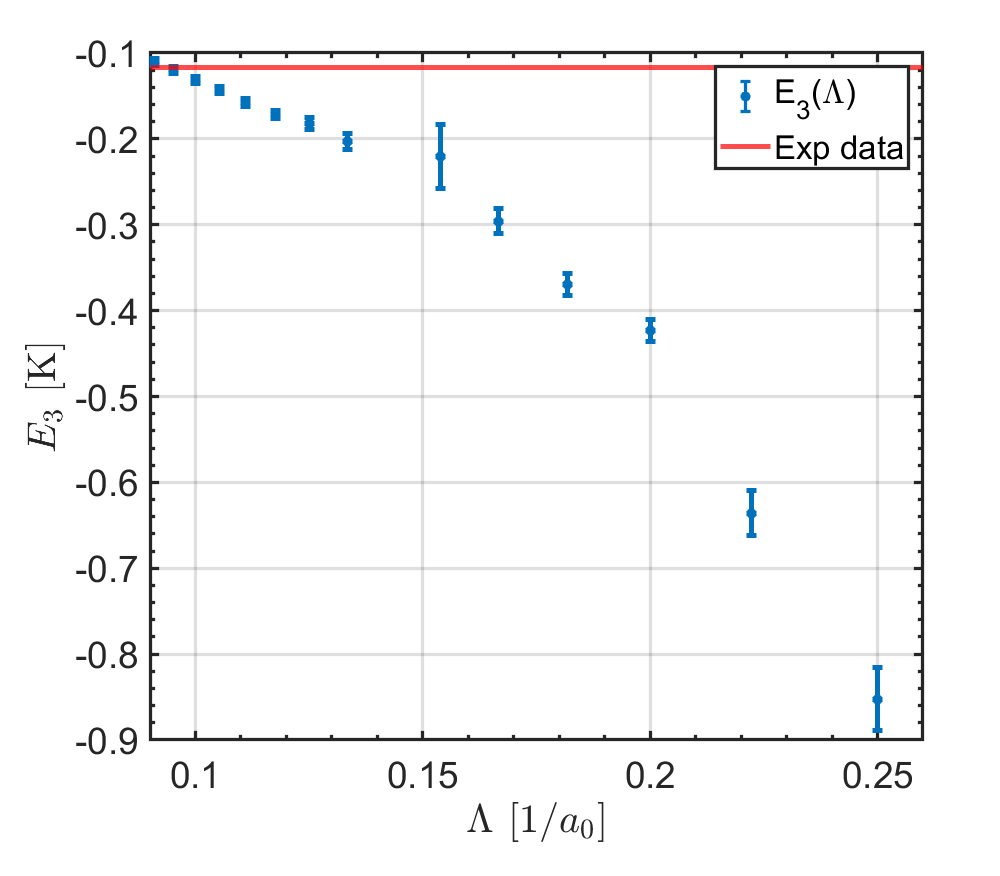}\
	\caption{}
	\end{subfigure}
		\begin{subfigure}[b]{0.235\textwidth}
	\includegraphics[width=1\linewidth]{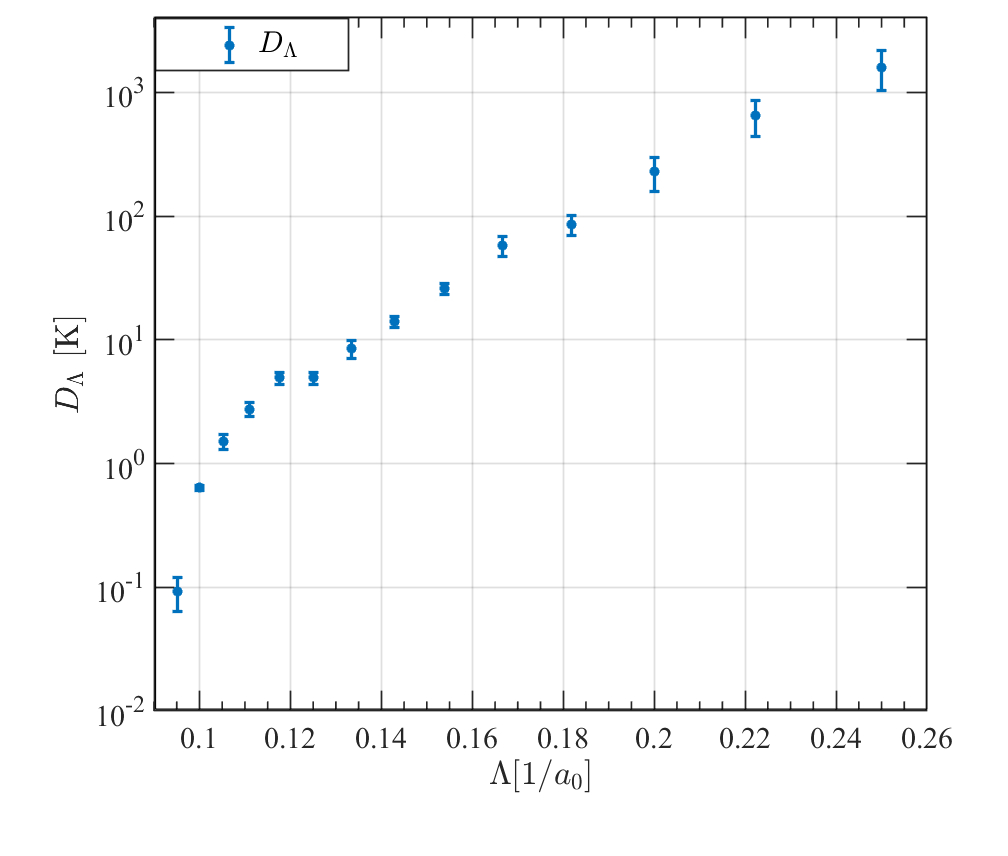}\
	\caption{}
	\end{subfigure}
	\caption{\footnotesize{The trimer binding energy ($E_3(\Lambda$) as a function of the cutoff $
	\Lambda$ (a) and the three-body force, $D(\Lambda)$ as a function of the cutoff $
	\Lambda$ (b)}}\label{fig_E3}
\end{figure} 
At leading order (LO), the numerical solution of the three-body Schrodinger equation for the trimer shows a strong cutoff dependence, i.e., $E_3=E_3(\Lambda)$ (see fig.~\ref{fig_E3}. This is expected, and it is a manifestation of the so-called Efimov effect~(see, e.g., \cite{3bosons, triton}), that leads to a collapse in the limit $\Lambda\rightarrow\infty$. Since the theory must be renormalizable,i.e., for each $\Lambda$, $E_3(\Lambda)=E_3$, to remove the cutoff dependence, one needs to add a three-body interaction and the corresponding LEC $D(\Lambda)$, such that:
\begin{eqnarray}\label{Eq_H3}
\hat{H}_N=-\sum_{i<j}\frac{\hbar^2}{2M_{\text{cm}}}\nabla_{ij}^2+V_2 (\Lambda)e^{-r_{ij}^2\cdot\Lambda^2}+\\
\nonumber
+\sum_{i<j<k}{D}_\Lambda(\Lambda)e^{-2\left(r_{ij}^2+r_{ik}^2+r_{jk}^2\right)^2\cdot\Lambda^2/3}~,
\end{eqnarray}
the trimer binding energy $D(\Lambda)$ is set using the HFDHE2 potential, which is $117.3$ mK (solid line in Fig.~\ref{fig_E3},a). Figure~\ref{fig_E3},b shows the resulting calibrated three-body force. 
For the case of the four-body system, it was already found in previous work, that no four-body counterterm is needed at LO \cite{Bazak:2016wxm}. Once the effective interaction is calibrated on the two- and three-body systems, we can test its applicability to the equation of state by first solving the problem for larger droplets as a consistency test, then extending the calculations to the bulk system.
		\begin{figure}[h]
		\includegraphics[width=1\linewidth]{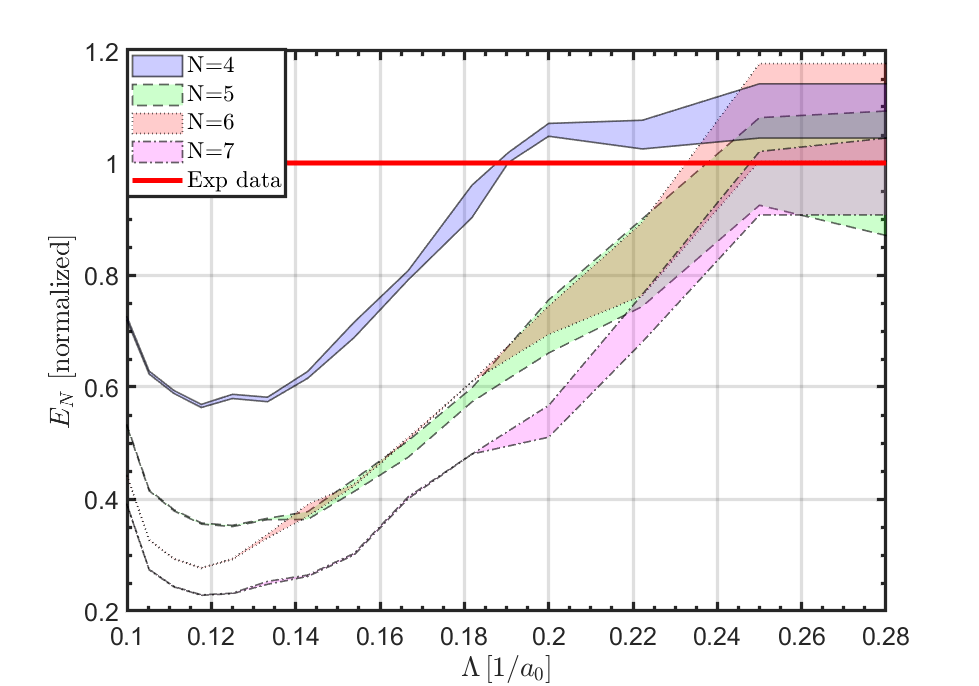}\\
\caption{DMC binding energies of N=$4-7$ $^4$He droplets described by a LO-effective contact potential as a function of the cutoff $
\Lambda$. The results are normalized to those obtained using the HFDHE2 potential~\cite{doi:10.1063/1.438007}). Blue solid band: $N=4$; Green dashed band: $N=5$; Red dashed band: $N=6$; Magenta dotted-dashed band: $N=7$. Solid red line: result from the HFDHE2 potential~\cite{doi:10.1063/1.438007}. The width of the bands reflects the statistical uncertainty intrinsic to the DMC algorithm.}\label{fig_4_7}
		\end{figure}

As mentioned, calculations were performed using the standard DMC algorithm, using a trial/importance function constructed as a Jastrow product of the analytic solutions $\Psi_{2-B}(\Lambda)$ of the two-body problem. In the cluster simulation the cutoff $\Lambda$ was taken in the range $0.1<\Lambda < 0.28 a_0^{-1}$.
Figure.\ref{fig_4_7} shows the DMC solution for the binding energy of $N=4-7$ systems of $^4$He droplets as a function of $\Lambda$. For $\Lambda>0.24$, for all the N-body systems, one obtains a reasonable convergence to energy close to the value predicted using the HFDHE2 potential. However, it is possible to see how the statistical uncertainty of the result greatly increases with the value of $\Lambda$. This is a signature of the increased fluctuations in energy due to the larger and larger steepness of the potential over a reduced spacial range. In any case we can consider the calibration of the effective interaction (that replaces a hard core two-body potential with a sum of two- and three-body potentials) as quite satisfactory.

\begin{figure}[h]
\includegraphics[width=1\linewidth]{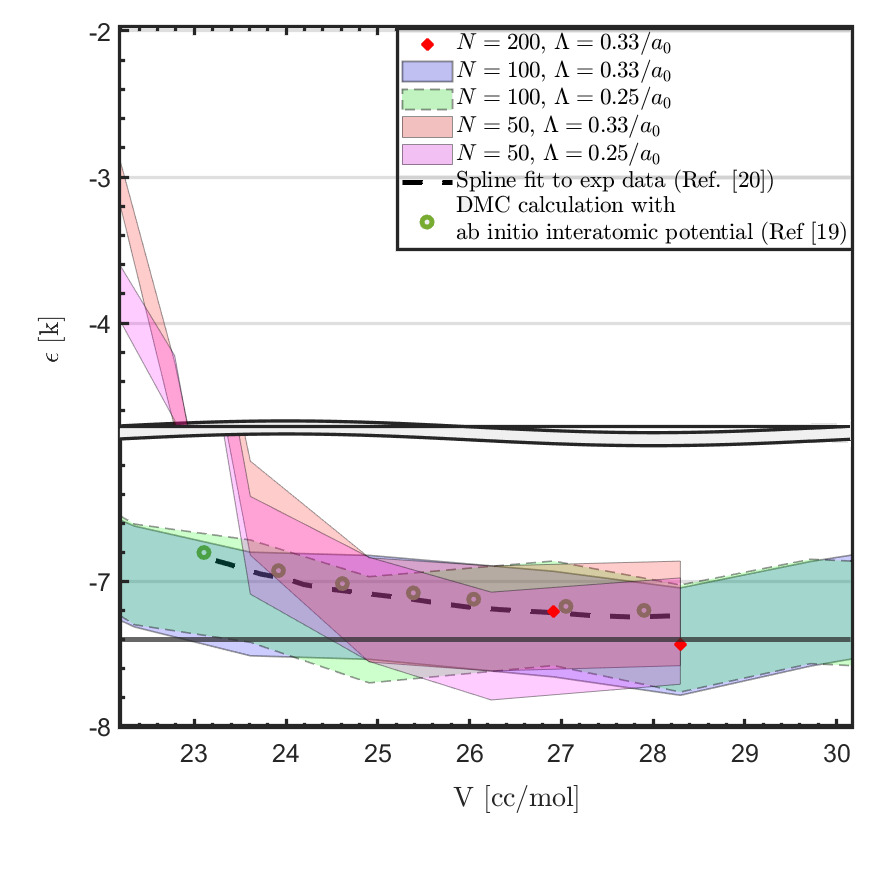}
\caption{DMC energy per  atom in $^4$He bulk liquid for periodic systems containing N=50,100 and 200 atoms and $\Lambda = 0.33/a_0$ and $0.25/a_0$. Blue solid band:$N=50$, $\Lambda=033/a_0$; Green solid band: $N=50$, $\Lambda=0.25/a_0$; Red Band: $N=100$, $\Lambda=0.33/a_0$; Magenta solid band: $N=100$, $\Lambda=0.25/a_0$; The red dots are  calculation for N=200 and $\Lambda=0.33/a_0$. Green circles: DMC calculation from Ref.~\cite{PhysRevLett.84.2650}. The dashed line reproduces the experimental values from ref.~\cite{ouboter1987thermodynamic}}\label{fig_condensed5}
		\end{figure}

The next step consists of extending the simulations to larger systems ($N = 50,100,200$), using the same two-body ansatz for the wavefunction, but imposing periodic boundary conditions. Under these conditions, one obtains a reasonable approximation of a bulk liquid. Calculations could be extended in principle to study the liquid/solid phase transition (see, e.g., \cite{PhysRevLett.84.2650}), but for the moment, we limit our analysis to a restricted interval around the saturation density, $27.55$ cc/mol \cite{PhysRevLett.84.2650}. It should be noticed that when simulating a bulk, the only relevant parameter should be the density $\rho=N/V$. 
In Figure \ref{fig_condensed5} we show the results of the DMC energy per particle of the bulk as a function of its specific volume $v=1/\rho$, expressed in cc/mol, for 5 different scenarios:
\begin{itemize}
 \item N=50, $\Lambda=033/a_0$ (Blue band)
 \item N=50, $\Lambda=0.25/a_0$ (Green band)
 \item N=100, $\Lambda=0.33/a_0$ (Red band)
 \item N=100, $\Lambda=0.255/a_0$ (Magenta band) 
 \item N=200 $\Lambda=0.33/a_0$ (Red dots)
\end{itemize}
For each of these $(N,\Lambda)$ pairs we calculated the equation of state $E(v)/N$ for a set of specific volumes around the experimental saturation value $v_0\simeq27.8$cc/mol, and compared to the experimental data from Ref.~\cite{ouboter1987thermodynamic} and the DMC calculation with a more accurate two- plus three-body interaction from Ref.~\cite{PhysRevLett.84.2650}. For both $N=50$ and $N=100$, the width of the band represents the statistical uncertainty, which is quite large due to the substantially large value of the cutoff employed. As it can be noted, energies are all compatible with the experimental and theoretical values. However, also the saturation density, despite the large uncertainty, encompasses the correct value. Interestingly, it can be noticed how for $N=50$ there is a very large deviation of the predicted energies at larger densities. This can be expected, since at such densities and for such a small number of particles, the size of the box becomes of order of twice the potential range, and hence the small box size enters an artificial artifacts to the Schrodinger equation, makes the calculation unreliable for these volumes. 

\begin{figure}[h]
\includegraphics[width=1\linewidth]{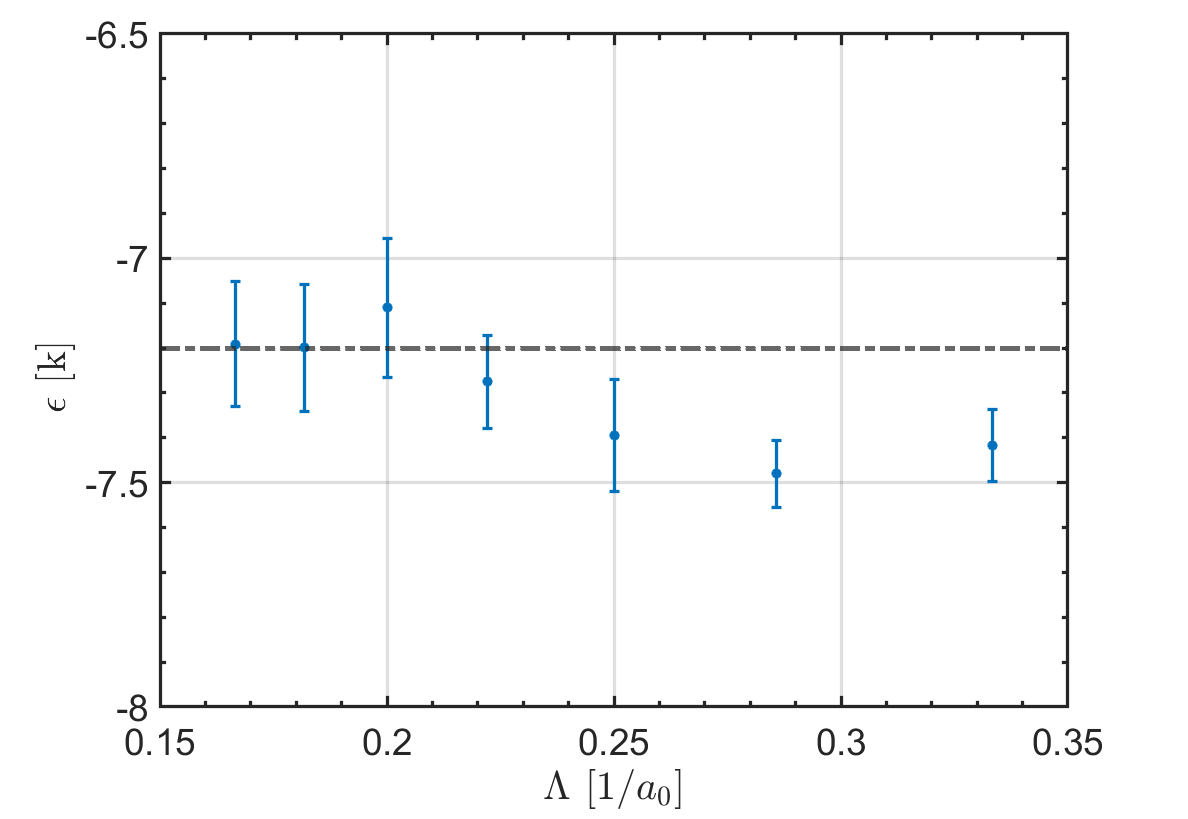}
\caption{Energy per particle $\epsilon$ at saturation volume $v=27.55$ cc/mol for $0.18/a_0\leq\Lambda\leq 0.33/a_0 $. Dashed line: experimental result $\epsilon= -7.17$ K \cite{ouboter1987thermodynamic} }\label{fig_multi_lambda}
\end{figure}

In order to better illustrate the cutoff dependence of the results, we computed in the liquid phase at the experimental saturation density the energy per particle for a periodic system with $N=100$ atoms using 8 different values of the cutoff $\Lambda$ in the range  $\Lambda=0.33/a_0$ to $\Lambda=0.18/a_0$. The results are shown in Fig.~\ref{fig_multi_lambda}. It can be seen that the numerical solutions display a mild cutoff dependency, In particular, for $\Lambda>0.25$ we can see a convergence towards an energy $\epsilon\simeq-7.4$ K, which is slightly deeper than the experimental value, $\epsilon=-7.17$ K  \cite{ouboter1987thermodynamic} and of the result of analogous DMC calculations in the bulk performed with the same HFDHE2 potential (see e.g. Ref.~\cite{PhysRevB.49.8920}). The 5\% difference can be attributed to the fact that this is a calculation at LO level. One can expect that introducing NLO corrections and beyond this gap can be reduced.

In any case, it should be noticed how the EoS estimated from the LO contact potential over a relatively wide range of specific volumes around the saturation points is qualitatively very well compatible with the physical behavior of the system. Since the energy can be in principle estimated correctly in a mean field calculation, this result is not very surprising. On the other hand, a correct value for the energy can be obtained in a model that does not respect more microscopic properties of the system. An interesting observable (that is related to neutron scattering experiments), is the so called pair distribution function $g(r)$, which is the Fourier transform of the static structure factor.
This is defined as (see e.g. Ref.\cite{PhysRevC.80.045802}):
\begin{equation}
 g(r) = \frac{1}{2\pi r\rho N}\times\sum_{i<j}\langle\psi|\delta\left(r_{ij}-r\right)|\psi_j\rangle~,
 \end{equation}
where $\rho$ is the density and $\rho g(r)d^3r$ is the probability of finding a $^4$He atom in an infinitesimal volume $d^3r$ at a distance $r$ from another $^4$He atom. The normalization is such that In the limit of large $r$, $g(r) \rightarrow 1$. 

The results for $g(r)$ as a function of the cutoff at saturation density are shown in Fig.~5. In contrast to what can be observed for the energy per particle $\epsilon$, where the dependence on $\Lambda$ is quite weak, the pair distribution function shows a completely different behavior. As we increase $\Lambda$, $g(r)$ displays a microscopic structure of the liquid that evolves from having a local cluster (reminiscent of Thomas collapse) to the typical structure of a liquid, with a peak at a distance slightly larger than the dimension of the atoms. The figure also reports $g(r)$ computed in a Variational Monte Carlo calculation using the HFDHE2 potential, which presents a hard core, and therefore a correlation hole that is much more pronounced. The striking result of our calculations is that increasing the cutoff one recovers almost quantitatively the microscopic features of the system, indicating that the effective interaction correctly describes the bulk at a very deep level.
		\begin{figure}[h]
\includegraphics[width=1\linewidth]{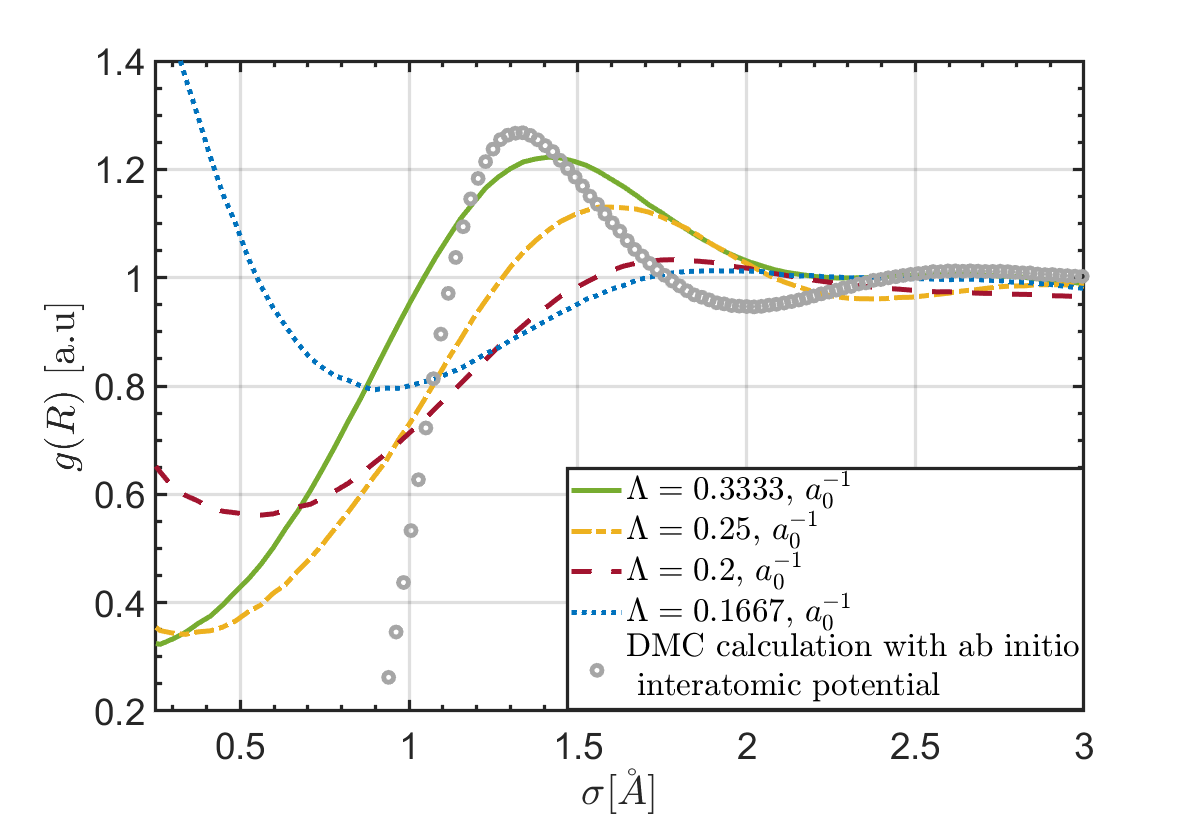}\\
\caption{Pair distribution function g(r) for different values of the cutoff $\Lambda$. Gray dots are VMC calculation with the HFDHE2 potential}\label{multi_lambda}
		\end{figure}
		\label{fig_g_r}
The results for $g(r)$ are strictly related to the evolution with the cutoff of the  effective two-body potential for the three body-system by integrating over $r_3$, defined as:
\begin{eqnarray}\label{eq_V2_eff}
 V_2^{\text{eff}}(\Lambda,r_{12})=V_2(\Lambda)\exp\left(-r_{12}^2/\Lambda^2\right)+
 \\
 \nonumber
 \int_0^{\infty}d^3r V_3(\Lambda)\exp\left[2\left(r_{12}^2+r_{23}^2+r_{13}^2\right)/3\Lambda^2\right]
 \end{eqnarray}
In Fig. 6 we report the DMC expectation of the effective potential in Eq.(\ref{eq_V2_eff}) for different values of the cutoff $\Lambda$. As it can be seen, the potential slowly develops a larger and larger repulsive part that eventually would well approximate the original hard core interaction.
		\begin{figure}[h]
			\centering
\includegraphics[width=1\linewidth]{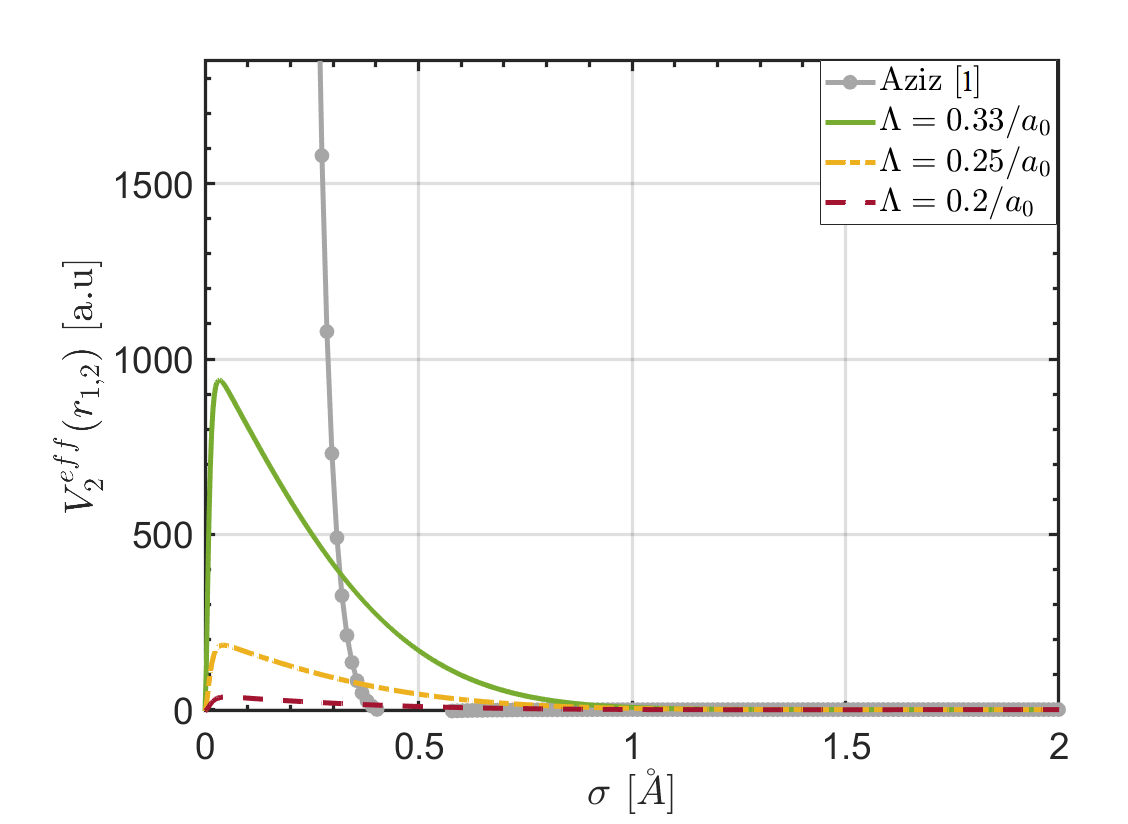}\\
\caption{Effective two-body interaction $V_2^{eff}(r_{12})$ as defined in Eq.(\ref{eq_V2_eff}) for different values of the cutoff  $\Lambda$. . The gray dots are the HFDHE2 potential by Aziz et al. \cite{doi:10.1063/1.438007}. }\label{fig_Aziz}
		\end{figure}
		\label{fig_g_r}

In conclusion, we have shown that approximating an hard core interaction in a many-Boson system (namely $^4$He atoms at T=0) with a contact interaction following the criteria that are commonly used to construct pionless effective potentials in nuclear physics (and first of all the renormalizability criterion), leads to a description of the bulk that for sufficiently large values of the cutoff in momentum space reproduce not only mean-field like quantities as the total enrgy, but also very sensitive microscopic observables such as  
 as the pair distribution function are seen to converge toward the exact results.
Hence, this work shows that even for bulks with relatively high density, perturbative EFT reproduces homogeneous matter's behavior and can be used for alternative solutions to exact potentials, especially when exact solutions are not possible. As a result, this work is an essential step toward calculating many-body systems and homogeneous nucleonic matter using an effective potential, removing the effects of fermionic nucleons and nuclear potentials that depend on relative spin and angular momentum.
	\begin{acknowledgments}
We thank Ubirajara van Kolck for important discussions which have contributed significantly 
to this work. H.D. was supported through a TIFPA-INFN/ECT* fellowship. All the numerical calculations were produced   
using HPC resources at CINECA through an INFN computing grant for the national theoretical initiative MONSTRE .
	\end{acknowledgments}
	\bibliography{references.bib}

\begin{thebibliography}{10}

\bibitem{doi:10.1063/1.438007}
R.~A. Aziz, V.~P.~S. Nain, J.~S. Carley, W.~L. Taylor, and G.~T. McConville.
\newblock An accurate intermolecular potential for helium.
\newblock {\em The Journal of Chemical Physics}, 70(9):4330--4342, 1979.

\bibitem{few}
Paulo~F. Bedaque and Ubirajara van Kolck.
\newblock {Effective field theory for few nucleon systems}.
\newblock {\em Ann. Rev. Nucl. Part. Sci.}, 52:339--396, 2002.

\bibitem{kaplan1}
David~B. Kaplan, Martin~J. Savage, and Mark~B. Wise.
\newblock {Nucleon - nucleon scattering from effective field theory}.
\newblock {\em Nucl. Phys.}, B478:629--659, 1996.

\bibitem{KSW1998_a}
Mark B.~Wise David B.~Kaplan, Martin J.~Savage.
\newblock {\em Nucl.Phys. B}, pages 329--355, 1998.

\bibitem{KSW1998_b}
Mark B.~Wise David B.~Kaplan, Martin J.~Savage.
\newblock {\em Phys.Lett. B}, pages 390--396, 1998.

\bibitem{KSW_c}
Mark B.~Wise David B.~Kaplan, Martin J.~Savage.
\newblock {\em Phys.Rev. C}, pages 617--629, 1999.

\bibitem{Bazak:2016wxm}
Betzalel Bazak, M.~Eliyahu, and U.~van Kolck.
\newblock {Effective Field Theory for Few-Boson Systems}.
\newblock {\em Phys. Rev. A}, 94(5):052502, 2016.

\bibitem{Platter:2004he}
L.~Platter, H.~W. Hammer, and Ulf-G. Meissner.
\newblock {The Four boson system with short range interactions}.
\newblock {\em Phys. Rev. A}, 70:052101, 2004.

\bibitem{PhysRevA.104.L030801}
A.~Kievsky, G.~Orlandini, and M.~Gattobigio.
\newblock Many-body energy density functional.
\newblock {\em Phys. Rev. A}, 104:L030801, Sep 2021.

\bibitem{PhysRevLett.50.1676}
V.~R. Pandharipande, J.~G. Zabolitzky, Steven~C. Pieper, R.~B. Wiringa, and
  U.~Helmbrecht.
\newblock Calculations of ground-state properties of liquid $^{4}\mathrm{He}$
  droplets.
\newblock {\em Phys. Rev. Lett.}, 50:1676--1679, May 1983.

\bibitem{Carlson:2014vla}
J.~Carlson, S.~Gandolfi, F.~Pederiva, Steven~C. Pieper, R.~Schiavilla, K.E.
  Schmidt, and R.B. Wiringa.
\newblock {Quantum Monte Carlo methods for nuclear physics}.
\newblock {\em Rev. Mod. Phys.}, 87:1067, 2015.

\bibitem{RevModPhys.73.33}
W.~M.~C. Foulkes, L.~Mitas, R.~J. Needs, and G.~Rajagopal.
\newblock Quantum monte carlo simulations of solids.
\newblock {\em Rev. Mod. Phys.}, 73:33--83, Jan 2001.

\bibitem{Kievsky:2017mjq}
A.~Kievsky, A.~Polls, B.~Juli\'a-D\'\i{}az, and N.K. Timofeyuk.
\newblock {Saturation properties of helium drops from a Leading Order
  description}.
\newblock {\em Phys. Rev. A}, 96(4):040501, 2017.

\bibitem{10.1093/comjnl/4.4.318}
L.~Fox.
\newblock {Chebyshev Methods for Ordinary Differential Equations}.
\newblock {\em The Computer Journal}, 4(4):318--331, 01 1962.

\bibitem{Barnea:2013uqa}
N.~Barnea, L.~Contessi, D.~Gazit, F.~Pederiva, and U.~van Kolck.
\newblock {Effective Field Theory for Lattice Nuclei}.
\newblock {\em Phys. Rev. Lett.}, 114(5):052501, 2015.

\bibitem{Kirscher:2015yda}
Johannes Kirscher, Nir Barnea, Doron Gazit, Francesco Pederiva, and Ubirajara
  van Kolck.
\newblock {Spectra and Scattering of Light Lattice Nuclei from Effective Field
  Theory}.
\newblock {\em Phys. Rev. C}, 92(5):054002, 2015.

\bibitem{3bosons}
Paulo~F. Bedaque, H.~W. Hammer, and U.~van Kolck.
\newblock {The Three boson system with short range interactions}.
\newblock {\em Nucl. Phys.}, A646:444--466, 1999.

\bibitem{triton}
Paulo~F. Bedaque, H.~W. Hammer, and U.~van Kolck.
\newblock {Effective theory of the triton}.
\newblock {\em Nucl. Phys.}, A676:357--370, 2000.

\bibitem{PhysRevLett.84.2650}
Saverio Moroni, Francesco Pederiva, Stefano Fantoni, and Massimo Boninsegni.
\newblock Equation of state of solid ${}^{3}\mathrm{He}$.
\newblock {\em Phys. Rev. Lett.}, 84:2650--2653, Mar 2000.

\bibitem{ouboter1987thermodynamic}
R~De~Bruyn Ouboter and Chen~Ning Yang.
\newblock The thermodynamic properties of liquid 3he-4he mixtures between 0 and
  20 atm in the limit of absolute zero temperature.
\newblock {\em Physica B+ C}, 144(2):127--144, 1987.

\bibitem{PhysRevB.49.8920}
J.~Boronat and J.~Casulleras.
\newblock Monte carlo analysis of an interatomic potential for he.
\newblock {\em Phys. Rev. B}, 49:8920--8930, Apr 1994.

\bibitem{PhysRevC.80.045802}
S.~Gandolfi, A.~Yu. Illarionov, F.~Pederiva, K.~E. Schmidt, and S.~Fantoni.
\newblock Equation of state of low-density neutron matter, and the
  ${}^{1}{S}_{0}$ pairing gap.
\newblock {\em Phys. Rev. C}, 80:045802, Oct 2009.

\end{thebibliography}
	\bibliographystyle{apsrev4-1}
	\bibliographystyle{unsrt}
\end{document}